\documentclass[12pt]{article}

\usepackage{graphicx}

\usepackage{scicite}

\usepackage{times}

\newenvironment{sciabstract}{%
\begin{quote} \bf}
{\end{quote}}

\newcounter{lastnote}
\newenvironment{scilastnote}{%
\setcounter{lastnote}{\value{enumiv}}%
\addtocounter{lastnote}{+1}%
\begin{list}%
{\arabic{lastnote}.}
{\setlength{\leftmargin}{.22in}}
{\setlength{\labelsep}{.5em}}}
{\end{list}}

\title{Discrete Hierarchical Organization\\of Social Group Sizes}

\author
{Wei-Xing Zhou,$^{1}$ Didier Sornette,$^{1,2,3\ast}$ Russell A. Hill,$^{4}$ Robin I.M. Dunbar$^{5\ast}$\\
\\
\normalsize{$^{1}$Institute of Geophysics and Planetary Physics,}\\%
\normalsize{University of California, Los Angeles, CA 90095, USA}\\
\normalsize{$^{2}$Department of Earth and Space Sciences,}\\%
\normalsize{University of California, Los Angeles, CA 90095, USA}\\
\normalsize{$^{3}$Laboratoire de Physique de la Mati\`ere
Condens\'ee,}\\
\normalsize{CNRS UMR 6622 and Universit\'e de Nice-Sophia
Antipolis,}\\
\normalsize{06108 Nice Cedex 2, France}\\
\normalsize{$^{4}$Evolutionary Anthropology Research Group,}\\
\normalsize{Department of Anthropology University of Durham, 43
Old Elvet,}\\
\normalsize{Durham DH1 3HN, UK}\\
\normalsize{$^{5}$British Academy Centenary Project, School of
Biological Sciences,}\\
\normalsize{University of Liverpool, Crown St., Liverpool
L69 7ZB, England}\\
\\
\normalsize{$^\ast$To whom correspondence should be addressed.}\\
\normalsize{E-mail: sornette@moho.ess.ucla.edu (D.S.) and
rimd@liverpool.ac.uk (R.I.M.D.).} }

\date{}

\begin{document}


\baselineskip24pt


\maketitle


\begin{sciabstract}
The ``social brain hypothesis'' for the evolution of large brains
in primates has led to evidence for the coevolution of neocortical
size and social group sizes. Extrapolation of these findings to
modern humans indicated that the equivalent group size for our
species should be approximately 150 (essentially the number of
people known personally as individuals). Here, we combine data on
human grouping in a comprehensive and systematic study. Using
fractal analysis, we identify with high statistical confidence a
discrete hierarchy of group sizes with a preferred scaling ratio
close to $3$: rather than a single or a continuous spectrum of
group sizes, humans spontaneously form groups of preferred sizes
organized in a geometrical series approximating $3, 9, 27,...$
Such discrete scale invariance (DSI) could be related to that
identified in signatures of herding behavior in financial markets
and might reflect a hierarchical processing of social nearness by
human brains.
\end{sciabstract}


Attempts to understand the grouping patterns of humans have a long
history in both sociology \cite{Coleman} and social anthropology
\cite{Kottak,Scupin}. However, these approaches have been largely
ecological in focus.  In contrast, recent attempts to understand
the evolution of sociality in primates have focussed in part on
the cognitive constraints that may limit the ecological
flexibility of group size \cite{D92JHE,D98EA}. The social brain
hypothesis, as it has come to be known, argues that the evolution
of primate brains (and in particular, the neocortex) was driven by
the need to coordinate and manage increasingly large social
groups.  Since the stability of these groupings is based on
intimate knowledge of other individuals and the ability to use
this knowledge to manage social relationships effectively, the
volume of neural matter available for cognitive processing
inevitably imposes a species-specific limit on group size.
Attempts to increase group size beyond this threshold result in
reduced social stability and, ultimately, group fission.
Extrapolating these findings to humans led to the prediction that
humans had a cognitive limit of about 150 on the number of
individuals with whom coherent personal relationships could be
maintained \cite{D93BBS}. Evidence to support this prediction has
come from a number of ethnographic and sociological sources. It
has, however, always been recognised that both human and nonhuman
primate groups are internally highly structured.  Further analyses
\cite{KudoDunbar} have indicated that at least one level of
structuring (the grooming clique) also correlates with neocortex
size. While it is not always clear what the significance of these
tiered groupings is, it is clear that human social groups (like
those of other primates) consist of a series of hierarchically
organised sub-groupings. We first review previous quantifications
of group sizes and then provide a systematic analysis. There is no
universally accepted procedure and all methods attempting of
identify group sizes suffer from several sources of bias (small
sample size, large inter-individual variability, and the criteria
used to include individuals). Our strategy is to include all the
reasonable data and attempt to extract useful signals above the
noise level by a careful analysis of the global data set.

The core social grouping is called the support clique, defined as
the set of individuals from whom the respondent would seek
personal advice or help in times of severe emotional and financial
distress, whose mean size is typically 3-5 individuals
\cite{DS95HN,HD03HN}. Above this may be discerned a grouping of
12-20 individuals (often referred to as a sympathy group) that
characteristically consists of all the individuals with whom one
has special ties; these individuals are typically contacted at
least once a month \cite{DS95HN,HD03HN}. The ethnographic data on
hunter-gatherer societies \cite{D93BBS} point to a grouping of
30-50 individuals as the size of overnight camps (sometimes
referred to as bands). These groupings are often unstable, but
their membership is always drawn from the same set of individuals,
who typically number in the order of 150 individuals. This last
grouping is often identified in small scale traditional societies
as the clan or regional group. Beyond these, at least two larger
scale groupings have been identified in the ethnographic
literature: the megaband of about 500 individuals and the tribe (a
linguistic unit, commonly of 1000-2000 individuals) \cite{D93BBS}.

\begin{figure}
\begin{center}
\includegraphics[width=8cm]{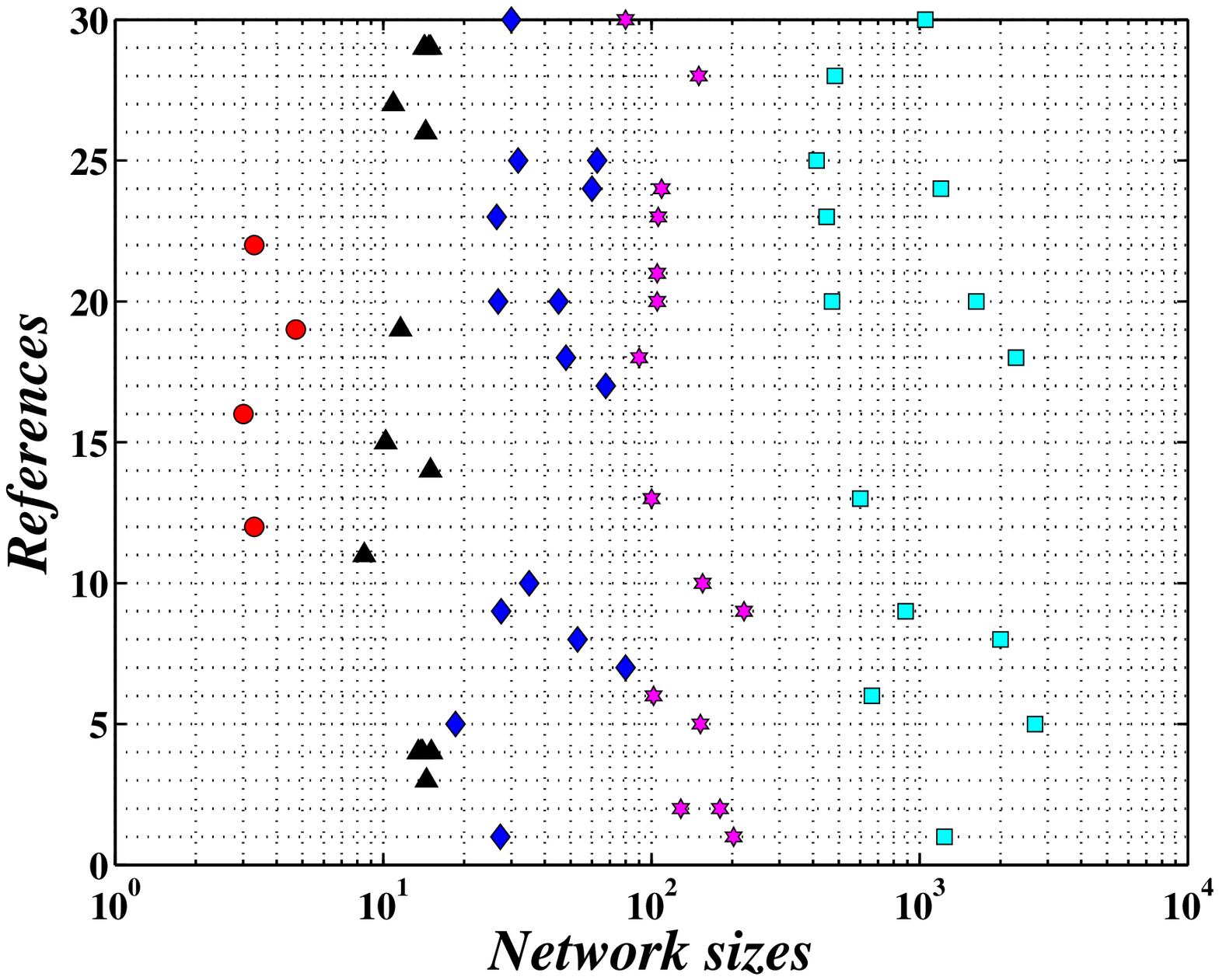}
\end{center}
\caption{Presentation of our data set of 61 group sizes. The
ordinate is an arbitrary ordering of data sources and the abscissa
gives the group sizes reported in each sources. The symbols refer
to the classification used in each of the studies: circle (support
clique), triangle (sympathy group), diamond (bands), stars
(cognitive groups), squares (small and large tribes). This
classification is not used in our systematic analysis summarized
in the other figures, to avoid any bias. } \label{Fig:NWDSI:Data}
\end{figure}

We complement the data used in Refs. \cite{D93BBS,DS95HN,HD03HN},
and sources therein with new data as follows. The USA 1998 General
Social Survey reports a mean size of 3.3 for support clique
\cite{M03SN}. The sizes of sympathy groups are reported to be 14.0
in Egypt, 15.1 in Malaysia, 13.5 in Mexico, 13.8 in South Africa
\cite{B92PR}, 10.2 in USA \cite{LMVKOC95SN}, 15 in The Netherlands
(1995) \cite{K97JVIB,KHH00SN}, 15.0 in The Netherlands (1992),
14.3 in The Netherlands (1992-1993), 14.8 in The Netherlands
(1995-1996), 14.2 in The Netherlands (1998-1999)
\cite{vTvG02JSI,vGvT03AS}, and 14.4 in Mali of West Africa
\cite{AMS02SSM}. See Figure \ref{Fig:NWDSI:Data}.

{\bf Method 1: Average sizes of different network layers}. To
summarize the previously cited data, we denote $S_1$ as the mean
support clique size, $S_2$ the mean sympathy group, $S_3$ the mean
band size, $S_4$ the mean cognitive group size, and $S_5$ and
$S_6$ the size of small and large tribes. Here, we do not address
the relevance of this classification (which will be done below)
but only characterize it quantitatively. The previously cited data
gives $S_0=1$ (individual or ego), $S_1=4.6$, $S_2=14.3$,
$S_3=42.6$, $S_4=132.5$, $S_5=566.6$, and $S_6=1728$. In order to
determine the possible existence of a discrete hierarchy, we
construct the series of ratios $S_i/S_{i-1}$ of successive mean
sizes: \begin{equation} S_i/S_{i-1} = 4.58, 3.12, 2.98, 3.11,
4.28, 3.05~,~~~ {\rm for}~~i = 1,\cdots,6~. \label{fkqllq}
\end{equation} This result suggests that humans form groups according to
a discrete hierarchy with a prefered scaling ratio between 3 and
4: the mean of $S_i/S_{i-1}$ is $3.50$.

\bigskip

{\bf Method 2: Probability density function and generalized
$q$-analysis of the complete data set.} In order to avoid any
potential biases in the published group classifications defined,
we employ a more systematic method of analysis that uses all the
available data and not just the mean group sizes.

\begin{figure}
\begin{center}
\includegraphics[width=8cm]{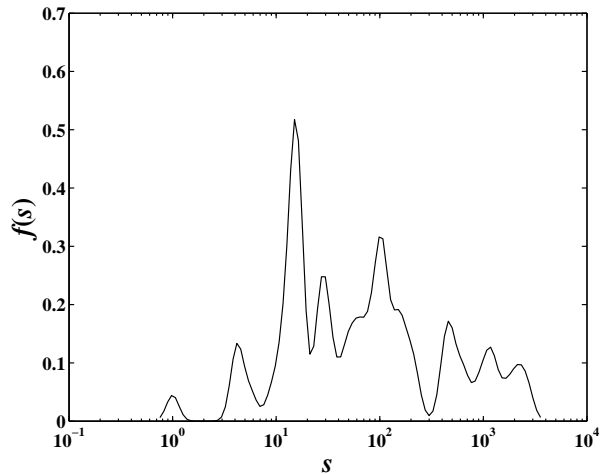}
\end{center}
\caption{Probability density function $f(s)$ of size $s$ estimated
with a Gaussian kernel estimator in the variable $\ln s$ with a
bandwidth $h=0.14$. Varying $h$ by 100\% does change $f(s)$ much
and gives similar results.} \label{Fig:NWDSI:fs}
\end{figure}

The sample has $61$ grouping clusters (including the ego) with
size $s_i$ available for $i=1,2,\cdots, 61$. Figure
\ref{Fig:NWDSI:Data} presents the data in a form attributing group
sizes to their relevant studies in an arbitrary order. We consider
this sample to be a realization of a distribution whose sample
estimation can be written as
\begin{equation}
 f(s) = \sum_{i=1}^{61}\delta(s-s_i)~,
 \label{Eq:Ns}
\end{equation}
where $\delta$ is Dirac's delta function. Figure
\ref{Fig:NWDSI:fs} shows the probability density function $f(s)$
obtained by applying a Gaussian kernel estimation approach
\cite{S86}.

Our challenge is to extract a possible periodicity in this
function in the $\ln s$ variable, if any. For instance, if the
ratios given in (\ref{fkqllq}) are genuine, one would expect a
periodic oscillation of $f(s)$ expressed in the variable $\ln s$
with mean period $\ln(3.5) =1.24$. This is called
``log-periodicity'' \cite{S98PR}.

\begin{figure}
\begin{center}
\includegraphics[width=8cm]{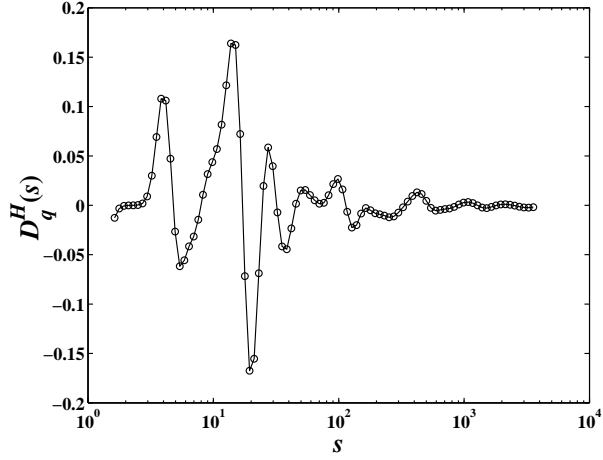}
\end{center}
\caption{Typical $(H,q)$-derivative $D_q^H(s)$ of the probability
density $f(s)$ as a function of size $s$ with $H=0.5$ and
$q=0.8$.} \label{Fig:NWDSI:Ds}
\end{figure}

Standard spectral analysis applied to $f(s)$ is dominated by the
trend seen in Figure \ref{Fig:NWDSI:fs} giving a peak at a very
low log-frequency corresponding to the whole range of the group
sizes. We thus turn to generalized $q$-analysis, or
$(H,q)$-analysis \cite{ZS02PRE}, which has been shown to be very
sensitive and efficient for such tasks. The $q$-analysis is a
natural tool to describe DSI in fractals and multifractals
\cite{E97PLA,EE97PRL}. The $(H,q)$-analysis consists in
constructing the $(H,q)$-derivative
\begin{equation}
D_q^H f(s) \stackrel{\triangle}{=} \frac
{f(s)-f(qs)}{[(1-q)s]^H}~. \label{Eq:HqD}
\end{equation}
Introducing an exponent $H$ different from $1$ allows us to
detrend $f(s)$ in an adaptive way. Note that the limit $H=1$ and
$q \to 1$ retrieves the standard definition of the derivative of
$f$. A value of $q$ strictly less than $1$ allows one to enhance
possible discrete scale structures in the data. To keep a good
resolution, we work with $0.65 \le q \le 0.95$, because smaller
$q$'s require more data for small $s$'s. To put more weight on the
small group sizes (which are probably more reliable since they are
obtained by conducting general surveys in larger representative
populations), we use $0.5\le H\le 0.9$. A typical
$(H,q)$-derivative with $H=0.5$ and $q=0.8$ is illustrated in a
semi-log plot in Figure \ref{Fig:NWDSI:Ds}.

\begin{figure}
\begin{center}
\includegraphics[width=8cm]{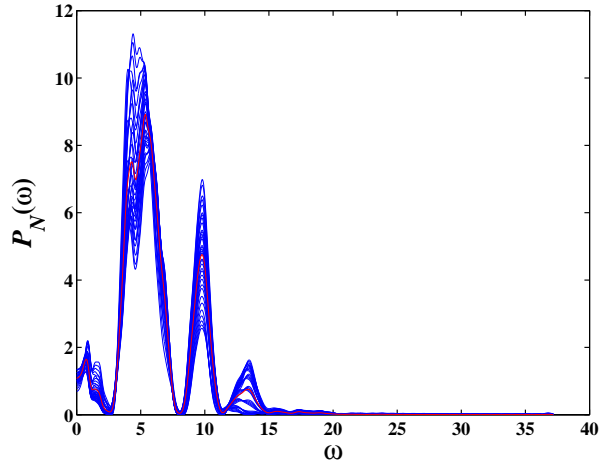}
\end{center}
\caption{Normalized Lomb periodograms $P_N(\omega)$ as a function
of angular log-frequency $\omega$ of the $(H,q)$-derivative
$D_q^H(s)$ for different pairs of $(H,q)$ with $0.5 \le H \le 0.9$
and $0.65 \le q \le 0.95$. The red line gives the averaged Lomb
power.} \label{Fig:NWDSI:LombAll}
\end{figure}

We then use a Lomb periodogram analysis \cite{Press} to extract
the log-periodicity in $f(s)$. Figure \ref{Fig:NWDSI:LombAll}
presents the normalized Lomb periodograms of $D_q^H f(s)$ for
different pairs of $(H,q)$ with $0.5 \le H \le 0.9$ and $0.65 \le
q \le 0.95$. This figure illustrates the robustness of our result.
For the specific values  $H=0.5$ and $q=0.8$ shown in Figure
\ref{Fig:NWDSI:Ds}, the highest peak is at $\omega_1 = 5.40$ with
height $P_N = 8.67$. The preferred scaling ratio is thus $\lambda
= {e}^{2\pi/\omega_1}=3.20$, which is consistent with the previous
result using the ``grouping analysis'' (\ref{fkqllq}). The
confidence level is 0.993 under the null hypothesis of white noise
\cite{Press}. If the underlying noise decorating the log-periodic
structure is correlated with a Hurst index of 0.6, the confidence
level decreases to 0.99; if the Hurst index is 0.7, the confidence
level falls to 0.85 \cite{ZS02IJMPC}.

The Lomb periodograms also exhibit a second peak at
$\omega_2=9.80$ with height $P_N=5.48$. This can be interpreted as
the second harmonic component $\omega_2 \approx 2 \omega_1$ of the
fundamental component at $\omega_1 = 5.40$. The amplitude ratio of
the fundamental and the harmonic is $1.26$. The co-existence of
the two peaks at $\omega_1$ and $\omega_2 \approx 2 \omega_1$
strengthens the statistical significance of a log-periodic
structure. To see this, we constructed 10000 synthetic sets of 61
values uniformly distributed in the variable $\ln s$ within the
interval $[0, \ln(2000)]$. By construction, these 10000 sets,
which are exactly of the same size as our data and span the same
interval, do not have log-periodicity and thus have no
characteristic sizes. We then applied the same procedure as for
the real data set to these synthetic data sets and obtain 10000
Lomb periodograms. We then performed the following tests on their
Lomb periodograms. Find the highest Lomb peak $(\omega, P_N)$. If
$P_N > 8.5$, check if there is at least another peak at $2\omega
\pm 1$ with its $P_N$ larger than $5.5$. 238 sets among the 10000
passed the test, suggesting a probability that our signal results
from chance equal to 0.024. The probability that there are at
least two peaks (one in $4.9<\omega<5.9$ with $P_N>8.5$ and the
other in $9.5<\omega<11.5$ with $P_N>5.5$) is found equal to
77/10000, giving another estimation of 0.993 for the statistical
confidence of our results. Another metric consists in quantifying
the area below the significant peaks found in the Lomb periodogram
of our data and comparing them with those in the synthetic sets.
We count the area of the main peak of the Lomb periodogram at
$\omega$ and add to it the areas of its harmonics whose local
maxima fall in the intervals $[(k-(1/5))\omega, (k+(1/5))\omega]$
for $k=2, 3, ...$ around all its harmonics. The area associated
with a peak is defined as the region around a local maximum
delimited by the two closest local minima bracketing it. The
fraction of synthetic sets which give an area thus defined larger
than the value found for the real data is 6-7\%, depending on the
specific values $H$ and $q$ used in the analysis. Summarising, all
these tests suggest that the evidence in support of our hypothesis
data is significantly unlikely to result from chance, but rather
reflects the fact that human group sizes are naturally structured
into a discrete hierarchy.

\bigskip

\begin{figure}
\begin{center}
\includegraphics[width=8cm]{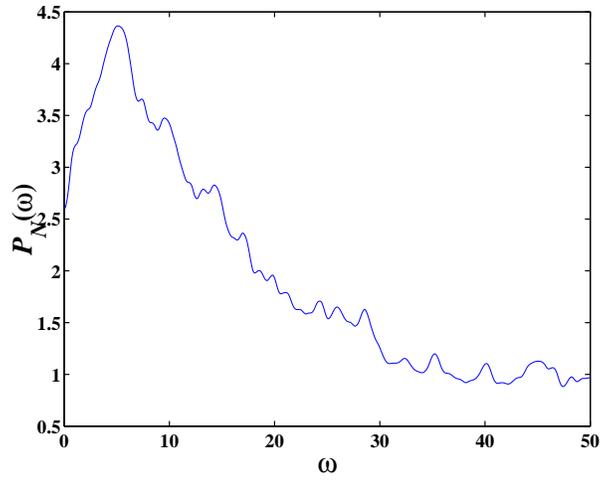}
\end{center}
\caption{Average Lomb periodogram $P_N(\omega)$ of the
$(H,q)$-derivative $D_q^H(s)$ with respect to the number of
receivers of the residual contact frequency for each individual in
the Christmas card experiment, as a function of the angular
log-frequency $\omega$ of the $(H,q)$-derivative, over the 42
individuals and different pairs of $(H,q)$ with $-1 \le H \le 1$
and $0.80 \le q \le 0.95$. } \label{Fig:NWDSI:XmasCard}
\end{figure}

{\bf Method 3: Probability density function and generalised
$q$-analysis of individual networks}.  We apply the same analysis
to individual social networks based upon the exchange of Christmas
cards in contemporary Western Society \cite{HD03HN}. This study
indicated that contemporary social networks may be differentiated
on the basis of frequency of contact between individuals, but that
both `passive' and `active' factors  may determine contact
frequency.  Controlling for the passive factors allowed the
hierarchical network structure to be examined on the basis of
residual (active) contact frequency.  Starting from the residual
contact frequencies, we constructed their $(H,q)$-derivative with
respect to the number of people contacted for each individual,
obtained the Lomb spectrum of the $(H,q)$-derivative and then
averaged them over the 42 individuals in the study (Figure
\ref{Fig:NWDSI:XmasCard}). The very strong peak at $\omega=5.2$ is
consistent with the previous results with a preferred scaling
ratio from the expression $\lambda = e^{2 \pi/\omega} \approx 3.3$
\cite{S98PR} for the smaller grouping levels in this study (group
sizes below 150).

\bigskip

{\bf Discussion}. Putting together a variety of measures collected
under a wide range of experimental conditions and in different
countries, we have documented a coherent set of characteristic
group sizes organized according to a geometric series with a
preferred scaling ratio close to $3$.
Similar hierarchies can be found in other types of human
organizations, of which the military probably provides the best
examples. In the land armies of many countries, one typically
finds sections (or squads) of about 10-12 soldiers, platoons (of 3
sections, $\approx 35$), companies (3-4 platoons, $\approx
120-150$), battalions (usually 3-4 companies plus support units,
$\propto 550-800$), regiments (or brigades) (usually three
battalions, plus support; $2500+$), divisions (usually 3
regiments), and corps (2-3 divisions). This gives a series with a
multiplying factor from one level to the next close to three.
Could it be that the army's structures have evolved so as to mimic
the natural hierarchical groupings of everyday social structures,
thereby optimising the cognitive processing of within-group
interactions?

The existence of a discrete hierarchy of group sizes may provide a
key ingredient in rationalizing the reported existence of discrete
scale invariance (DSI) in financial time series in so-called
``bubble'' regimes characterized by strong herding behaviors
between investors \cite{mybookcrash}. Johansen et al.
\cite{Ledoit1,Ledoit2} have proposed a model to explain the
observed DSI in stock market prices as resulting from a discrete
hierarchy in the interactions between investors. Recent analysis
of DSI in regimes with strong herding component have also
identified the presence of a strong harmonic at $2\omega$, similar
to the findings reported here \cite{JS99IJMPC,SZ02QF}. The fact
that DSI is found only during stock market regimes associated with
a strong herding behavior suggests that it may reflect the fact
that a discrete hierarchy of naturally occurring group sizes
characterizes human interactions whether they be hunter-gatherers
or traders. The present work suggests that this discrete hierarchy
may have its origins in the fundamental organization of any social
structure and be deeply rooted within the cognitive processing
abilities of human brains.

When dealing with discrete hierarchies, it may be important to
distinguish between the specific group sizes on the one hand and
their successive ratios on the other. It may be that the absolute
values of the group sizes are less important than the ratios
between successive group sizes. If the ratio of group sizes is
interpreted as a fractal dimension (specifically, the ratio is
related to the imaginary part of a fractal dimension: see
\cite{S98PR} and references therein), this would imply that,
depending on the social context, the minimum ``nucleation'' size
may vary, but the ratio (close to 3) might be universal. The
fundamental question, then, is to determine the origin of this
discrete hierarchy. At present, there is no obvious reason why a
ratio of 3 should be important. Equally, however, we have little
real understanding of what cognitive mechanisms might limit the
nucleation point to a particular value. Considerable additional
work will need to be done on both these components if we are to
understand why these constraints on human grouping patterns exist
and exactly what their significance might be.


\begin{scilastnote}
\item Research by WXZ and DS was partially supported by the James
S. Mc Donnell Foundation 21st century scientist award/studying
complex system. Research by RH and RD was funded by the ESRC's
Research Centre in Economic Learning and Social Evolution (ELSE).
RD's research is supported by the British Academy Centenary
Project and by a British Academy Research Professorship.
\end{scilastnote}

\clearpage

\clearpage

\end{document}